\documentstyle[12pt]{article}
\def\beq{\begin{equation}}
\def\eeq{\end{equation}}
\def\bea{\begin{eqnarray}}
\def\eea{\end{eqnarray}}
\def\bq{\begin{quote}}
\def\eq{\end{quote}}

\def\beqa{\begin{eqnarray}}
\def\eeqa{\end{eqnarray}}
\def\be{\begin{equation}}
\def\ee{\end{equation}}
\def\beq{\begin{equation}}
\def\eeq{\end{equation}}

\def\pa{\partial}
\def\kaps{{\kappa}^{2}}

\def\bi{\begin{itemize}}
\def\ei{\end{itemize}}

\def\ov{\overline}

\def\lc{{\cal L}}

\begin{document}
\pagestyle{empty}
\begin{flushright}
CERN-TH/99-185\\ 
hep-th/9906148
\end{flushright}
\vspace*{1cm}
\begin{center}
{\Huge Supergravity and Supersymmetry Breaking in Four and Five Dimensions
} \\ \vspace*{5mm} \vspace*{0.5cm} John
Ellis$^{a)}$, Zygmunt Lalak$^{b,c)}$, Stefan Pokorski$^{c)}$ and
Steven Thomas$^{d)}$\\ \vspace{0.3cm} \vspace*{1cm} {\bf
ABSTRACT}\\
\end{center}
\vspace*{5mm} \noindent
We discuss supersymmetry breaking in the field-theoretical
limit of the strongly-coupled heterotic string compactified
on a Calabi-Yau manifold, from the different
perspectives of four and five dimensions. The former applies
to light degrees of freedom  
below the threshold for five-dimensional
Kaluza-Klein excitations, whereas the five-dimensional
perspective is also valid up to the Calabi-Yau scale. We show how, in
the latter case, two gauge sectors separated in the fifth dimension
are combined to form
a consistent four-dimensional supergravity. In the lowest order of the 
$\kappa^{2/3}$ expansion, we show how a
four-dimensional supergravity with gauge kinetic function
$f_{1,2}=S$ is reproduced, and we show  
how higher-order terms give rise to four-dimensional operators
that differ in the two gauge sectors.  
In the four-dimensional approach, supersymmetry is seen to be broken 
when 
condensates form on one or both walls, and the goldstino may have a
non-zero
dilatino component. As in the five-dimensional approach,
the Lagrangian is not a perfect square, and we have not identified
a vacuum with broken supersymmetry and zero vacuum energy.
We derive soft supersymmetry-breaking terms for non-standard
perturbative embeddings, that are relevant in more general
situations such as type I/type IIB orientifold models.\\

\vspace*{0.5cm} \noindent

\rule[.1in]{16.5cm}{.002in}

\noindent $^{a)}$ Theory Division, CERN, Geneva, Switzerland\\
$^{b)}$ Physikalisches Institut, Universit\"at Bonn, Germany\\
$^{c)}$ Institute of Theoretical Physics, Warsaw University,
Poland\\ $^{d)}$ Department of Physics, Queen Mary and Westfield
College, London UK\\ \vspace*{0.5cm}
\begin{flushleft}
June 1999
\end{flushleft}
\vfill\eject

\setcounter{page}{1} \pagestyle{plain}

\section*{Introduction}

In their impressive analysis of the effective field theory
limit of the strongly-coupled heterotic $E_8 \times E_8$  string
theory, Horava and Witten~\cite{witten,wh} constructed a consistent
eleven-dimensional supergravity on a manifold $M_4 \times X \times
S^1 / Z_2$, coupled to ten-dimensional Yang-Mills models
on the fixed hyperplanes of the $S^1 / Z_2$ orbifold.
Witten~\cite{witten} also solved the equations of motion 
along the eleventh dimension on the
orbifold $S^1 / Z_2$, and found the correct six-dimensional
compactification that preserves four unbroken supercharges in the
presence of non-trivial background components of the
antisymmetric tensor field $G_{ABCD}$. He also calculated the
gravitational constant in four dimensions and the gauge couplings
on both the visible and hidden walls.

On the basis of general arguments,
this should reduce to some four-dimensional 
$N = 1$ supergravity theory
in the infrared limit, but the above
results were obtained without constructing explicitly the
effective Lagrangian in four dimensions. Knowing that the final
effective theory is four-dimensional $N=1$ supergravity,  one way
to obtain the complete Lagrangian is simply to read the
K\"ahler potential, superpotential and gauge kinetic functions
off from the direct $d = 11 \rightarrow 4$ reduction of the relevant
eleven-dimensional terms. With the full four-dimensional Lagrangian at
hand, one can study the properties of its vacuum, such as supersymmetry
breaking, and thereby understand the infrared limit of
the strongly-coupled heterotic superstring.

We follow this route in the first part of this paper,
and contrast
the generic features of supersymmetry breaking due to gaugino
condensation~\cite{nilles1} in the strongly-coupled case~\cite{laltom}
with the
better-known weakly-coupled case~\cite{gweak}.
An important difference is that supersymmetry is
generically broken by non-zero expectation values of both $F^S$ and
$F^T$ in the strongly-coupled string. 
The effective potential of the four-dimensional
supergravity in the strongly-coupled case 
cannot be written simply as a perfect square, complicating
the minimization problem, which is not solved simply by
minimizing the $|F^S|^2$ term alone. 
This observation may be welcome, given the phenomenological
interest~\cite{louis} in the dilaton-dominated 
scenario for supersymmetry breaking.
However, we also demonstrate that
the potential does not vanish generically.

The four-dimensional approach is appropriate if the condensation scale
$\Lambda$ is smaller than the threshold $m_5 = 1/R_5$ for
five-dimensional Kaluza-Klein excitations.
However, studying the compactification chain  
$d = 11 \rightarrow 5 \rightarrow 
4$~\cite{bd,peskin,ovr5,elpp,elp,stelle} is justified
whatever the scale $\Lambda$ 
of the nonperturbative physics.
This approach allows us to study how 
the low-energy four-dimensional model arises out of the two 
$d = 10 \rightarrow 4$ sectors
which are spatially disconnected in the original 
$d = 11 \rightarrow 5$ theory, in other
words, how the two gauge sectors are glued together in
course of the compactification process.

It was noted in the earliest days of M phenomenology that
the size of the eleventh (fifth) dimension must be larger than the
Calabi-Yau radius, if one enforces unification of the gauge 
and gravitational couplings~\cite{witten}.
Therefore, it is physically more
interesting to compactify first the six Calabi-Yau dimensions:
$d = 11 \rightarrow 5$, and only
later the fifth dimension of the resulting five-dimensional
supergravity: $ d = 5 \rightarrow 4$. It has also been stressed~\cite{stelle}
that the $d = 11 \rightarrow 5$ compactification is, in
the presence of a non-trivial background for $G$, mathematically
more consistent than the path $d = 11 \rightarrow 10 \rightarrow 4$.
It can be thought of as a Calabi-Yau compactification with
non-vanishing values of the four-form field $G_{ABCD}$ in the
internal Calabi-Yau directions~\cite{stelle,strom,talk,dfkz}, i.e.,
on a manifold with torsion. Such a
construction gives gauged five-dimensional supergravity, as was
first emphasized in~\cite{ovr5}, see also~\cite{elp},
whose equations of motions have
domain-wall solutions. These
may be viewed as creating spontaneously the $S^1 /Z_2$
orbifold and reproduce the Witten solution
in the approximation linear in $x^{11}$.
The domain wall which is the vacuum
solution of the gauged supergravity is a BPS state, i.e., it
breaks four out of the eight supercharges possessed by the simple
ungauged supergravity/Abelian Yang-Mills theory in five
dimensions.

The complete compactification of that five-dimensional model down
to four dimensions has not yet been performed. An important aspect
of the attempt to establish the detailed relation
between the Horava-Witten model~\cite{wh} and the four-dimensional
supergravity is that
the Horava-Witten Lagrangian, although anomaly-free and
supersymmetric, is only a limited approximation to
the strongly-coupled string. It contains terms
which are of the order $\kappa^{2/3}$ with respect to the
gravitational action, plus those terms of order $\kappa^{4/3}$
which are necessary  to maintain  supersymmetry in eleven dimensions.

However, when one integrates out the fifth dimension to obtain various
terms in the four-dimensional Lagrangian, 
such as two- and four-fermion
terms, one finds that those portions of these terms that  differ
from one wall to the other arise at order $\kappa^{4/3}$ and
higher. Thus, in order to reproduce them reliably, one would
need to go beyond the linearized solution given by Witten~\cite{witten},
even
beyond the original eleven-dimensional field-theoretical Lagrangian. The
problem is underlined by the observation  that supersymmetry in
various dimensions interrelates terms that are of
different orders in the expansion parameter
$\kappa^{2/3}$. As we shall argue below, one important implication
of this situation is that certain features of the Horava-Witten
approximation to the strongly-coupled heterotic dynamics, such as
the global supersymmetry-breaking mechanism of Horava~\cite{hg}, may be
artefacts of the approximation, that are not present in the
four-dimensional
effective supergravity. It is nevertheless instructive to compactify
explicitly the fifth dimension and see the
structure of the four-dimensional supergravity in the consistent lowest-order
approximation in $\kappa^{2/3}$.

One issue to be discussed within this framework is
the transmission of supersymmetry breaking by the
five-dimensional bulk 
supergravity~\cite{elpp,elp,peskin}~\footnote{Discussions of
alternative approaches to supersymmetry breaking, e.g.,
the Scherk-Schwarz mechanism, averaging over
the fifth dimension, anomaly mediation and invoking a tower of
Kaluza-Klein
states can be found in~\cite{emilian,nom,sun,mno}. These have
features that differ from the mechanism discussed here.}.
The five-dimensional supergravity
formulation is particularly suitable for studying the
dynamical supersymmetry breaking of the residual four supercharges
and its transmission between spatially-separated gauge sectors,
for the following reasons.
Since we know completely the structure of the simplest
five-dimensional supergravity, we can
study the breaking of supercharges
within a fully-consistent supersymmetric model. Secondly, 
we can identify already in five dimensions
all the massless degrees of freedom of the effective
four-dimensional supergravity model. The bulk fields have well-defined
properties under the $Z_2$ orbifold parity, and therefore one
can write down $Z_2$-invariant couplings with the
walls. As we see in more detail later, 
the supersymmetry breaking on the visible
wall occurs locally,
without any need to average over the fifth dimension, 
once some fields or
their derivatives acquire vev's that are non-vanishing 
locally on the wall, as a result of equations of motion
in the fifth dimension.

In the present letter, we complete the previous work first by adding
explicitly to the bulk Lagrangian studied in~\cite{elpp} the most relevant
Lagrangian terms on the four-dimensional walls, as obtained by dimensional
reduction of the original Horava-Witten wall Lagrangian from $d = 10
\rightarrow 4$ on a Calabi-Yau manifold.  Next, we reduce the fifth dimension
by solving the equations of motion and integrating out the components of the
five-dimensional fields which depend on the fifth coordinate. The equations
of motion are determined by the structure of the five-dimensional theory, and
it turns out that we can solve them explicitly only in the linearized
approximation. In this way, we identify the five-dimensional origins of
various parts of the four-dimensional effective action, in particular those
that play a role in gaugino condensation.  As already mentioned, this
explicit construction is complete only to lowest order in $\kappa^{2/3}$ and,
therefore, the four-dimensional supergravity obtained in this approximation
is different from that considered in Section 2, which includes fully the
thershold corrections to the gauge kinetic functions that arise in the
strongly-coupled heterotic $E_{8} \times E_{8}$ string.  Finally, we address
the subtle question how the nonlinearities appearing in the solutions of the
five-dimensional equations of motion for the bulk moduli fields fit into the
canonical structure of the four-dimensional supergravity. 


\section*{Supersymmetry Breaking in Four Dimensions}

We begin by following the first route described in the Introduction.
We recall that one obtains from the Horava-Witten model~\cite{wh} 
the K\"ahler functions
and the gauge kinetic functions for the 
standard and non-standard embeddings,
consistently to order $\kappa^{2/3}$, 
that is with the threshold corrections
to the gauge kinetic functions included. 
This is sufficient to reconstruct directly
the parts of the scalar potential that are relevant
for seeing the supersymmetry-breaking structure in the 
effective four-dimensional supergravity theory
arising from the strongly-coupled heterotic string
below the scale $m_5 =
\frac{1}{R_5}$. In this case, it is fully adequate 
to work entirely within the four-dimensional supergravity
framework, as we assume in this Section.

We first
recall the way the vev of gaugino bilinears $\langle
\lambda^a  \lambda^b \rangle $ enters the 
effective scalar potential of the
four-dimensional supergravity~\cite{nfer}~\footnote{We
denote by $\lambda^a $ the gaugino components, where $a$ is an 
adjoint group index, and $i$ labels complex moduli fields.}.
Using the canonical normalization in four dimensions for
the gravitational, gauge and gaugino kinetic terms, the relevant part of
the Lagrangian is 
\beq
V  = \; e^K g^{i \bar{j}} (D_i W + \frac{1}{4} e^{-K/2} \partial_i
f_{ab} \langle \lambda^a \lambda^b \rangle ) (D_{\bar{j}} W +
\frac{1}{4} e^{-K/2} \partial_{\bar{j}} \bar{f}_{ab}
 \langle \bar{\lambda}^a
\bar{\lambda}^b  \rangle )+ \; ..., \label{gbil} 
\eeq 
where $g^{i
\bar{j}}$ is the inverse K\"ahler metric and rest of the notation
is standard~\cite{nfer}.

Comparing (\ref{gbil}) with the well-known 
general expression 
$V = g_{i \bar{j}} F^i F^{\bar{j}} - 3 e^G$ 
for the four-dimensional potential in
terms of the $F^i$, the auxiliary fields for the chiral multiplets,
we read off the
modified expressions for the auxiliary fields in the presence of
the condensates 
\beq 
\label{efs} 
F^i=e^{K/2} g^{i \bar{j}}
(D_{\bar{j}} \bar{W} + \frac{1}{4} e^{-K/2} \partial_{\bar{j}}
\bar{f}_{ab} \langle \bar{\lambda}^a \bar{\lambda}^b \rangle ).
\eeq
In the following, we match this expression
explicitly to the fermionic bilinears in the effective Lagrangian,
since~\cite{laltom} this is a better description when the
gauge kinetic function depends on more than one modulus, 
and the gaugino bilinears are among the
terms obtainable directly from the Calabi-Yau reduction.

We recall briefly the situation in the weakly-coupled
heterotic string. There, at tree level, the gauge kinetic function
is universal: $f=S$, and the K\"ahler function for the illustrative
case of a single universal
modulus $T$ is $K= - \log (S + \bar{S}) - 3 \log (T + \bar{T}
)$. In this case, the full potential
\beq 
V = g_{S \bar{S}} | F^S |^2 + g_{T
\bar{T} } |F^T|^2 - 3 |W|^2 e^K 
\eeq 
reduces to 
\beq V = g_{S
\bar{S}} | F^S |^2 , 
\eeq 
since $g_{T \bar{T} } |F^T|^2 = 3 |W|^2
e^K$. This relation is equivalent to the vanishing of the perfect square
containing the gaugino condensates in the ten-dimensional
effective action of the weakly-coupled heterotic superstring
\cite{ibanez}. In this way, we find
\beq
F^S = 0, \;\; F^T \neq 0
\label{fslessft}
\eeq
At the level of quantum corrections, there are additional
contributions to the gauge kinetic functions which depend on
the modulus $T$, and additional terms required by the space-time
modular invariance appear in terms containing gaugino bilinears,
and hence in the effective superpotential describing condensation.
However, in all the models studied so far, the vacuum relation $F^S
\ll F^T$ persists, and supersymmetry breaking occurs along the
$T$ direction in the moduli space. The technical reason
is that the dependence on $S$ factorizes out in simple modular-invariant
models of condensation, and the equations of motion,
i.e., dynamics, tell us that $F^S$ is small. 
The vev of the modulus $T$
is rather small in these models, close to unity in supergravity units.

The structure of the supersymmetry-breaking sector is significantly
modified in the strongly-coupled
regime.
The K\"ahler function for the universal moduli $S,\, T$
is the same, but the classical, or tree level, gauge kinetic functions
are changed, and are different for different walls: $f_{1,2} = S
\pm \xi_0 T $  where 
\beqa
\label{e:xpl} \xi_0 &=& - \frac{ \pi \rho_0 }{2 (4 \pi)^{4/3}
\kappa^{2/3}} \frac{1}{8 \pi^2} \int_X \omega_{K} \, \wedge \, (\,
trF^{(1)} \wedge F^{(1)}
 - \frac{1}{2} \, trR \wedge R \,).
\eeqa 
Here $\omega_K$ is the K\"ahler $(1,1)$-form, and the
topological integral over Calabi-Yau space can be parametrized in terms
of gauge and gravitational instanton numbers characterizing the
embedding~\cite{lpt}:
\beq 
\xi_0 = \frac{n_{F1} - \frac{1}{2} n_{R}}{32
\pi^3}, 
\label{xizero}
\eeq 
The interesting region of moduli space is
where $S={\cal O}(2)$ and $T={\cal O}(80)$~\cite{lpt,sstieb}. 
Hence, we are not
interested in mechanisms which generate minima of the potential at
$T \approx 1$, but need some new mechanism which generates a
minimum in the region of current interest. We do not
discuss any specific mechanism here, but just state the
possibilities opened up by the current form of the kinetic functions.

First, we note that $S$ and $T$ enter the kinetic functions, and
hence any nonperturbative  potential, in quite a symmetric way.
The relative coefficient $\xi_0$ (\ref{xizero}) which weights the
contribution of
$T$ changes from model to model. In the elliptic-fibration
models of~\cite{sstieb}, this number is smaller than
$0.025$. It could in principle be either much larger or much smaller
in more general constructions.
Because of thegreater symmetry between $S$ and $T$,
there is no obvious reason why 
$F^S$ should be much smaller than $F^T$ in the
generic case.
In the strongly-coupled case,
we obtain
\begin{eqnarray} 
& V = e^K (S + \bar{S})^2 \left | -\frac{W}{S + \bar{S}} +
\frac{1}{4} e^{-K/2} ( \Lambda^{3}_1 + \Lambda^{3}_2) \right |^2 &
\nonumber \\ & + e^K \frac{(T + \bar{T})^2}{3}  \left |
-\frac{3W}{T + \bar{T}} + \frac{1}{4} \xi_0 e^{-K/2} (
\Lambda^{3}_1 - \Lambda^{3}_2) \right |^2 - 3 e^K \left | W \right
|^2. & 
\end{eqnarray} 
and the result for the $F$ terms is~\footnote{Here we suppress
the possibility of a constant superpotential contribution, which
could arise as a vev of the gauge and/or
gravitational Chern-Simons forms on either wall, whose inclusion
does not change the general picture.}:
 \beq
 F^S = \frac{1}{4} (S + \bar{S} )^2 ( \Lambda^{3}_1 + \Lambda^{3}_2 )
 \eeq
 \beq
 F^T = \frac{1}{12} (T + \bar{T} )^2 \xi_0 ( \Lambda^{3}_1 - \Lambda^{3}_2 )
 \eeq
It is clear that supersymmetry is unbroken: $F^S = F^T = 0$
if and only if both
condensates vanish. Even if there is only one
condensate, both $F^S$ and $F^T$ are non-zero.
Moreover,
if condensates on both walls
are switched on simultaneously, no matter in what proportion,
supersymmetry is always broken in four dimensions. 
In particular,
even when the two condensates are switched on with
the same magnitude, and opposite signs, supersymmetry is formally
broken~\footnote{The magnitude of the breaking is to be
determined from the vacuum solution of the effective potential,
but we would expect $m_{3/2} = {\cal O}(\Lambda^3/M^2_{P})$.},
contrary to~\cite{hg}.
A further consequence of $F^S \ne 0$ is a non-zero scalar
mass, which arises from (\ref{gbil}) upon
substituting the correction
$- \delta K_S = \pm \xi_0 |A^q |^2 /
(S + \bar{S})^2 $~\footnote{This correction arises from the
correction to the metric of the Calabi-Yau space, i.e., to the factor
of $\sqrt{g_{(6)}}$ that multiplies the kinetic terms of the 
four-dimensional charged scalars \cite{ovr4}.}, 
which gives soft scalar masses
proportional to $ \pm \xi_0 F^S$.

Finally, we examine the ratio of the two $F$ terms 
\beq
\frac{F^S}{F^T} = \frac{3}{\xi_0} \frac{ \Lambda^{3}_1 +
\Lambda^{3}_2 }{\Lambda^{3}_1 -  \Lambda^{3}_2 } \left
(\frac{S}{T} \right )^2. 
\eeq 
In the present region of moduli
space, the ratio of $S/T$ is of the order $1/40$ or so, so 
it would not require very much fine-tuning to arrange the magnitudes of
the condensates
in such a way that the ratio $F^S/F^T$ is of the order of unity or
larger. To make the possibility of the mixed
$S,T$-moduli-driven scenario more plausible, we look at the
ratio $F^S/F^T$ more carefully. As pointed out in~\cite{lpt}, one
can easily express the expectation value of $T$ through the
observable quantities $T=(\frac{M_P}{M_{GUT}})^2
\frac{\alpha_{GUT}^{1/3}}{ 2^{17/3} \pi^3 } $. Then we can express
$S$ as $S=\frac{1}{4 \pi \alpha_{GUT}} - \xi_0 T$.
As a result, we obtain the ratio of the $F$ terms as a
function of $\xi_0$: 
\beq 
\frac{F^S}{F^T} = \frac{3}{\xi_0} \left (
\frac{2^{11/3} \pi^2 M_{GUT}^{2}}{\alpha_{GUT}^{4/3} M_{P}^2 }
 - \xi_0 \right )^2 \frac{ \Lambda^{3}_1 + \Lambda^{3}_2 }{\Lambda^{3}_1 -
\Lambda^{3}_2 }. 
\eeq 
The prefactor multiplying the condensates can be
studied as a function of $\xi_0$, when we fix the observables at
their MSSM values. One finds that the prefactor vanishes at
${\xi_0} \approx 0.025$, but grows quickly to values of
${\cal O}(1)$ for larger $\xi_0$, and to the values $ \ge 0.07$ at
$\xi_0 \le 0.01$. For
negative $\xi_0$, i.e., in the regime of `strong'
unification, the value of the prefactor is always larger than
$1/10$. Thus, it is possible to obtain quite a large value of
$F^S$, and even the extreme option of
$S$-dilaton-driven supersymmetry breaking cannot be
excluded in the strongly-coupled heterotic string.
This could have interesting consequences, given the promising results
of phenomenological investigations of this limit 
in the weakly-coupled string. 

We finish this section with a list of the soft
termswe found in the four-dimensional supergravity
approach, which is the way the soft terms were obtained in
all the papers published so far:~\cite{anqu,nom,st,choi,kob}.
We generalize the earlier
results by considering non-standard embeddings
in which charged matter is present on both walls, i.e., in both
gauge sectors, and we allow for condensates to form on both
walls~\footnote{However, we do not consider five-branes in the bulk.}. 
Restoring powers of the reduced Planck mass $M$, 
the physical gravitino mass 
in four dimensions is:
\beq 
m^{2}_{3/2} = \frac{(S +
\bar{S})^{2/3}}{2^{2/3} 12 M^4} (\Lambda_{1}^{3} + \Lambda_{2}^{3}
)^2 + \frac{(T + \bar{T})^2}{ 432 M^4} \xi^{2}_{0} (
\Lambda^{3}_{1} - \Lambda^{3}_{2} )^2 
\eeq 
and the mixing angle
$\theta$ introduced through the relation  $
\frac{F^S}{F^T} = \sqrt{3} \frac{S + \bar{S}}{T + \bar{T} } \tan
\theta$ is given by 
\beq 
\tan \theta  = \sqrt{3} \xi^{-1}_{0}
\frac{S + \bar{S}}{T + \bar{T} } \frac{\Lambda_{1}^3 +
\Lambda_{2}^3 }{\Lambda_{1}^3 - \Lambda_{2}^3 }. 
\eeq 
Assuming that the
CP-violating phases vanish, we obtain trilinear scalar
terms of the form 
\beq 
\label{tri} A = \sqrt{3} m_{3/2} \left (
\sin \theta \left ( -1 \pm \xi_0 \frac{3 ( T + \bar{T} )}{3 ( S
+ \bar{S} ) \pm \xi_0 ( T + \bar{T})} \right ) + \sqrt{3} \cos
\theta  \left ( -1 +\frac{3 ( T + \bar{T} )}{3 ( S + \bar{S} )
\pm \xi_0 ( T + \bar{T})} \right ) \right ) \eeq and gaugino
masses \beq \label{gauginos} M_{1/2} = \frac{\sqrt{3} m_{3/2} }{
(S + \bar{S} ) \pm \xi_0 ( T + \bar{T} ) } \left ( \sin \theta
(S + \bar{S}) \pm \xi_0 \cos \theta  \frac{(T + \bar{T}
)}{\sqrt{3}} \right ). 
\eeq 
Note that there is a
difference of sign between the expressions linear in $\xi_0$
corresponding to different walls. This can have
consequences in some of non-standard embedding models, where, e.g.,
matter with Standard Model hypercharge may
exist on both walls. The dilaton-dominated limit corresponds to
$\sin\theta \rightarrow 1$. 
We can see from the formulae for the $A$ terms
and gaugino masses that even in this limit there is
non-universality between terms containing charged fields
from different walls.
As for soft scalar masses in this context,
one can easily imagine strongly-coupled heterotic string
models analogous to weakly-coupled orbifolds with twisted matter,
where the scalar masses for many twisted fields are non-zero. Using
standard supergravity formulae, one can easily write down the
expressions for these masses, in terms of the 
corresponding K\"ahler potential $K$,
\beqa
& m^{2}_{i \bar{j}} = K_{i \bar{j}} \left ( m^{2}_{3/2} - 
\frac{3 m^{2}_{3/2} }{3 ( S + \bar{S} ) \pm \xi_0 (T + \bar{T}) }
 \left ( \pm \xi_0  ( T + \bar{T}) (2 -
\frac{\pm \xi_0 ( T + \bar{T} ) }{3 ( S + \bar{S} ) \pm  \xi_0 
( T + \bar{T} ) } ) \sin^2 \theta  \right . \right . & \nonumber \\
& \left . \left . + (S + \bar{S}) ( 2 - \frac{ 3 ( S + \bar{S} ) }{3 ( S + \bar{S} ) 
\pm  \xi_0 ( T + \bar{T} ) } ) \cos^2 \theta  - \frac{ \pm \xi_0 2 \sqrt{3} (T + \bar{T}) 
(S + \bar{S})}{3 ( S + \bar{S} ) \pm  \xi_0 
( T + \bar{T} ) } \sin \theta \cos \theta \right ) \right ).&
\eeqa
One again notices the
characteristic differences of sign between terms from different
wall, and the universality of these terms is violated
even in the dilaton-driven supersymmetry-breaking limit.


\section*{The Five-Dimensional Connection between Four-Dimensional Worlds}

In this Section, we construct explicitly the four-dimensional supergravity
Lagrangian that arises from the sequence of compactifications: $d = 11
\rightarrow 5 \rightarrow
4$, remembering that the result can be trusted only in the lowest
order in $\kappa^{2/3}$. In this case, the four-dimensional Lagrangian is born
from two spatially-separated
gauge sectors. As will be clear from our construction,
the result obtained at the lowest order of the expansion in $\kappa^{2/3}$
should be a four-dimensional supergravity theory in the
approximation $f_{1,2}=S$.

The general five-dimensional model obtainable through
compactification of eleven-dimensional supergravity on the
deformed Calabi-Yau space constructed in~\cite{witten} has
already been given in~\cite{elp}. There were worked out the couplings of
some important
operators belonging to the Yang-Mills sectors living on the walls,
such as gaugino bilinears and scalar trilinear operators derived
from the Lorentz Chern-Simons terms, to scalar fields living in
the five-dimensional bulk. Here we concentrate on details of
these couplings which are relevant for the
construction of the effective four-dimensional model, and identifying
the operators that violate four-dimensional supersymmetry.

We must first return to the Bianchi identity for
$G_{11ABC}$,
and to the perfect-square structure of Horava~\cite{hg}.
Horava has shown that, for the Lagrangian describing the
eleven-dimensional
supergravity coupled to two Yang-Mills sectors on the
ten-dimensional boundaries to be supersymmetric, 
one needs among other terms a
specific pair of gaugino interaction terms:
\begin{eqnarray}
& \sum_{i=1,2}
\left
( \frac{\sqrt{2} }{96 \pi (4 \pi \kappa^2 )^{2/3}} \int_{M^{(i)}}
d^{10} x \sqrt{g} \bar{\chi^a} \Gamma^{ABC} \chi^a G_{ABC11} \right . &
\nonumber \\
 &- \left .
\frac{\delta ( x^{5}_{(i)} ) }{96 (4 \pi )^{10/3} \kappa^{2/3} }
\int_{M^{(i)}} d^{10} x \sqrt{g} \bar{\chi^a} \Gamma^{ABC} \chi^a
\bar{\chi^b} \Gamma_{ABC} \chi^b \; \right ) 
\label{nterms}&
\end{eqnarray}
where $i=1,2$ labels the walls and $x^{5}_{(1)} = 0, \;
x^{5}_{(2)} = \pi \rho_0 $. We write these terms in this form to
stress that they should be interpreted as boundary terms.
Indeed, the first, non-singular term is simply the well-known
coupling
of the ten-dimensional gauginos to the antisymmetric tensor field.
Since the product $ \delta (x^{11} ) \, \delta
(x^{11} - \pi \rho_0 ) $ vanishes, 
the combination of terms involving $G_{11ABC}$ and ten-dimensional
gauginos $ \chi $ (\ref{nterms}) can be written as a {\em bulk}
perfect-square action: 
\beq
\label{eq:10} 
L_{sq}  =  - \frac{1}{12
\kaps} \int_{M^{11}} d^{11} x \, \sqrt{g} \, ( G_{11ABC} - \sum_m
\frac{\sqrt{2}}{16 \pi} {(\frac{\kappa}{4 \pi})}^{2/3}
\delta^{(m)} ( x^{11} )\, {\rm Tr} \, \bar{\chi}^{(m)}
\Gamma_{ABC} \chi^{(m)} \,  )^2, 
\eeq 
where
\bea 
&G_{11abc} = (\partial_{11} C_{abc} \pm 23 \; perms) +
{\kappa^2 \over \sqrt{2} \lambda^2} \delta(x^{11}) ({2 \over 3} tr
A_a[A_b,A_c] +cycl.) & \nonumber \\ & + {\kappa^2 \over \sqrt{2}
\lambda^2} \delta(x^{11}-\pi \rho_0) ({2 \over 3} tr A_a[A_b,A_c]
+cycl.).& 
\eea 
We have retained in this expression only these parts of the
gauge Chern-Simons forms which give rise to couplings between zero
modes arising from the compactification of the six-dimensional
internal space, and are directly related to the low-energy
effective superpotential.

Since, in this Section, we first construct the 
five-dimensional theory, 
we perform the Weyl rotation of the metric
which gives the canonical Einstein-Hilbert action in five dimensions.
The relation between the canonical eleven- and five-dimensional 
metrics is
$g^{(11)}_{MN} = (e^{-2 \sigma } g^{(5)}_{\mu \nu}\, ; \, e^\sigma
g^{(0)}_{ab} ) $. In this notation, we take $e^{3 \sigma} = \,
Re(S)$, where $S$ is the $Z_2$-even scalar from the universal
hypermultiplet~\cite{elpp,elp}. This five-dimensional moduli
$S, T, U$ should not be confused with the four-dimensional chiral
superfields
encountered in the previous Section, which we denote by
$\tilde{S}, \tilde{T}, \tilde{U}$ 
when necessary in this Section.
In what follows, we shall use the same symbols both for 
other five-dimensional moduli, e.g.,
$\sigma, \gamma$, and for the corresponding four-dimensional quantum
fields,
whenever it is obvious from the context which we actually
have in mind. We need the decompositions of eleven-dimensional 
spinors $\chi_L
(x,y)= \chi_L (x) \otimes \eta_+ (y)$, where $\eta_+$ is the
covariantly-constant spinor of positive chirality~\footnote{We choose the
basis in which negative-chirality spinors on the walls are projected
out by $Z_2$ parity.}. The corresponding decomposition of $\Gamma$
matrices is $ \Gamma^{M}_{(11)} = \{\gamma^{m}_{(11)} \otimes {\bf
1}\,;\, i \bar{\gamma} \otimes \gamma^k \} $ in tangent indices,
where $\bar{\gamma} \equiv i \gamma^0 \gamma^1 \gamma^2 \gamma^3$,
$\Omega \equiv \gamma^7=i \prod_k \gamma^k$, $\bar{\gamma} \Omega=
\Gamma^{11}$. The relation between $\gamma$ matrices with world
indices in eleven- and five-dimensional normalizations is
$\Gamma_{(11)}^{\mu} =
e^\sigma \gamma^{\mu}_{(5)} \, , \Gamma_{(11)}^{M} = e^{-\sigma/2}
\Gamma^{M}_{(0)} $, where the subscript $(0)$ denotes the fiducial
metric on the zeroth-order (undeformed) Calabi-Yau manifold.

If we restrict ourselves to a Calabi-Yau space with $h_{(2,1)}=0$, then
the decomposition of gauge fields with compact indices which
defines matter fields in ${\bf 27}$s of $E_6$ is $ A_i = A^{kp}(x)
\omega_{i}^{kj}(y) T_{jp} $, where the $\omega_{i}^{kj}(y)$ are 
harmonic $(1,1)$-forms, and the $T_{jp}$ are generators of $E_6$. We note
the following properties of the generators: $ {\rm Tr} T_{ip} T_{jq}
T_{kr} = \epsilon_{ijk} d_{pqr}$, $ {\rm Tr} T_{jp} T_{j'p'} =
\delta_{jj'} \delta_{pp'} $, and the appropriate expansions for the
field strength: $ F_{\mu a} \rightarrow \partial_{\mu} C^{Kp}
\omega_{a}^{Kj} T_{jp} $, $ F_{\nu b} \rightarrow \partial_{\nu}
C^{Lp'} \omega_{b}^{Lj'} T_{jp'}$. Finally, we shall use the
following decomposition of the regular part of $G_{abc11}$ in
terms of  massless Calabi-Yau modes~\cite{elp}
\be
(G_{11})_{abc} = 2 \partial_{11} C_0 \ov{\Omega}_{abc} +h.c. 
\ee
With these conventions and definitions, the result of the reduction
of the perfect square down to five dimensions is 
\beq 
\label{sq5} 
{\cal
L}_{sq} = - \frac{V_0}{12 \kaps} \int d^5 x \sqrt{g_{(5)}} e^{-3
\sigma} g^{55} \left ( 2 \partial_5 C_0 + 4
\frac{\kappa^2}{\sqrt{2} \lambda^2} \delta^{(m)} P^{(m)} - \frac{
1}{32 \pi } \left ( \frac{\kappa}{4 \pi} \right )^{2/3}
e^{\frac{9}{2} \sigma } (g_{55})^{3/4} \delta^{(m)}
\bar{\epsilon}^{(4)}_{m} \epsilon^{(4)}_{m} \right )^2 
\eeq 
where
$P^{(m)} (A) = \lambda_{KLM} d_{pqr} A^{Kp} A^{Lq} A^{Mr}$,
$m=1,2$ labels the walls, and we have used the canonical normalization
for gauginos in four dimensions. Here $e^{3 \sigma(x,x^5)}$ is the 
five-dimensional variable measuring the volume of the Calabi-Yau space
along the orbifold interval, in units of the fiducial volume $V_0$.
This is the real part of the $Z_2$-even scalar from the universal
hypermultiplet: $Re(S)= e^{3 \sigma (x^5)}$. The relation between
$S$ and the four-dimensional fields $\tilde{S},\, \tilde{T}$ is
\beq 
\label{rel45} 
e^{3 \sigma(x^5)} = e^{3 \sigma } + \xi_0
e^{\gamma} (1 - \frac{2 x^5}{\pi \rho_0} ).
\eeq
In the above
expression, $e^{3 \sigma} = Re (\tilde{S})$, and $e^{\gamma} = Re
(\tilde{T}) = \sqrt{g_{55} (x^\mu )}$ are functions of the four 
noncompact coordinates.

To
construct the effective theory in four dimensions, one has to
integrate out the components of the five-dimensional fields which do not
correspond to massless degrees of freedom in
four dimensions, and the natural
way to do this is through the solution of the equations of motion
along the dimension which one wishes to compactify.
To the lowest order in $\kappa^{2/3}$,
i.e., to the zeroth order in $\xi_0$, the equation is 
\beq 
\label{sol}
\partial^{2}_5 C_0 = \sum_m \left (- \frac{\sqrt{2} \kaps }{\lambda^2}
P^{(m)} + \frac{1}{64 \pi } \left( \frac{\kappa}{4 \pi} \right
)^{2/3} e^{\frac{9}{2} \sigma } (g_{55})^{3/4}
\bar{\epsilon}^{(4)}_m \epsilon^{(4)}_m \right ) \,
\partial_5 \delta^{(m)} 
\eeq
The solution to this equation which obeys the periodicity condition on
the full circle and is antisymmetric across the fixed points of the
$S^1 / Z_2 $  has a finite discontinuity at each of these points. Its
derivative develops $\delta$-function singularities at $x^5 =0,
\pi \rho_0 $, which cancel other $\delta$-function terms coming from the
expansion of the formal `square'. The regular part of the derivative, 
which is continuous everywhere, is 
\beq 
\label{p5c} 
\partial_5
C_0 = \frac{1}{2 \pi \rho_0 } \sum_{m=1,2} \left (- \frac{\sqrt{2}
\kaps }{\lambda^2} P^{(m)} + \frac{1}{64 \pi } \left(
\frac{\kappa}{4 \pi} \right )^{2/3} e^{\frac{9}{2} \sigma_m }
(g_{55\, m})^{3/4} \bar{\epsilon}^{(4)}_m \epsilon^{(4)}_m \right
). 
\eeq 
where the subscript $m=1,2$ denotes the restriction of the given
function to the $m$'th wall. We note that the coefficients of the
gaugino bilinears  above would differ in higher order in $\xi_0$
(see (\ref{rel45})). The effective four-dimensional Lagrangian is obtained
by substituting (\ref{p5c}) into Lagrangian (\ref{sq5}) and then
integrating over the coordinate $x^5$.

We need to comment on
the next-order corrections to the above solution. As we stressed
earlier, higher-order corrections to (\ref{p5c}) cannot be
reliably calculated from the Lagrangian (\ref{sq5}). It turns out
that, in order to find the corrections reliably and to reconstruct
the complete four-dimensional Lagrangian at higher order in $\xi$, it
would be necessary to go beyond the linear
order in the $x^5$ dependence of the field $C_0$,
hence beyond the order to which the
effective Lagrangian can be trusted. Secondly, the
compactification $11 \rightarrow 5$ leads in general to a nonlinear
$\sigma$-model structure which goes beyond the simple expression
(\ref{sq5}): as pointed out in~\cite{elp}, in five dimensions the 
perfect square  (\ref{sq5}) is a part of the
larger nonlinear $\sigma$-model structure 
\beq 
\label{sigm5}
L_{coupling} = -\frac{1}{2}g_{xy} g^{55} (\pa_5 \sigma^x - \lc
\delta (x^5 - \pi \rho) \delta^{x x_0}) (\pa_5 \sigma^y - \lc
\delta (x^5 - \pi \rho) \delta^{y x_0}),
\ee 
where we
assume that the four-dimensional gaugino
condensate is proportional to the Calabi-Yau $(3,0)$-form
$\Omega_{ijk}$. The operators $\lc$ contain the terms trilinear in
matter fields and bilinear in gauginos. The scalars $\sigma^x$ are
components of the five-dimensional hypermultiplets,
including the universal
one: $(S,C_0);(Z,C_1);...$. The approximate expression for the metric
is 
\beq 
\label{metric} g = \left[
\begin{array}{cccc}
{1 \over (S + \bar{S})^2} &  {{-2(C_0 +\bar{C_0})} \over  (S +
\bar{S})^2} & 0 &  {{-2(C_1 +\bar{C_1})} \over  (S + \bar{S})^2}
\\
 {{-2(C_0 +\bar{C_0})} \over  (S + \bar{S})^2} & {2 \over S + \bar{S}} &
 {{2(C_1 +\bar{C_1})} \over  S + \bar{S}} &  {{2(Z +\bar{Z})} \over  S
+ \bar{S}}
\\
0 &  {{2(C_1 +\bar{C_1})} \over  S + \bar{S}} & 1 &  {{2(C_0
+\bar{C_0})} \over  S + \bar{S}} \\
 {{-2(C_1 +\bar{C_1})} \over  (S + \bar{S})^2} &  {{2(Z +\bar{Z})} \over
 S + \bar{S}} &  {{2(C_0 +\bar{C_0})} \over  S + \bar{S}} &
{2 \over S + \bar{S}}
\end{array}
\right], 
\eeq 
when we expand the quaternionic metric up to linear
order in all the fields except $S$. There exist additional sources
for the field $S$ on the walls, coming from the additional 
ten-dimensional terms
which can be read from the supersymmetrization of the ten-dimensional
Green-Schwarz terms: $F^{2}_{(i)} \rightarrow F^{2}_{(i)} -
\frac{1}{2} R_{mn} R^{mn}$~\cite{romans}.
These terms are the dominant source of the vacuum domain-wall
configuration, but they do not contain observable fields. Hence, as
far as the matter fields and gaugino couplings are concerned,
the relevant couplings to moduli are in (\ref{sigm5}).
We reiterate that
nonlinearities are to be expected in the solution of the
full theory in five dimensions, since it contains nonlinear terms
associated with gauging, as well as with the nonlinear $\sigma$
model structure.

It is useful to summarize certain properties of the
zeroth-order result (\ref{p5c}). First, we recall~\cite{elpp,elp} that the
the order parameter for supersymmetry breaking in
the microscopic five-dimensional vacuum is the non-vanishing
vev of $\partial_5 C_0$ 
if we compactify on a Calabi-Yau space with $h_{(2,1)}=0$: see 
 (\ref{repl}) below for the general case with $h_{(2,1)} > 0$.
Secondly, (\ref{p5c}) expresses this quantity
in terms of fields charged under the gauge group, which
are legitimate zero modes in the effective four-dimensional
theory. We see that the terms corresponding to
superpotentials generated on separate walls enter this expression
additively, which is simply the minimal structure for
building the effective four-dimensional supergravity. This is not
surprising, knowing the 
way the parts of the superpotential add up in the effective Lagrangian
derived from the weakly-coupled string, 
but the low-energy superpotential could in principle be a
non-trivial chiral function of $W^1$ and $W^2$.
Thirdly, the terms
corresponding to gaugino bilinears enter the expression (\ref{p5c})
with coefficients that are equal to the order
to which we have solved the corresponding equation of motion.
We shall now discuss some of these points in more detail.


In the five-dimensional approach which we assumed in this Section,
following~\cite{elpp,elp}, it is straightforward to identify the
terms that violate the four low-energy supersymmetries,
namely, the terms containing couplings of the charged fields to the
derivative $\partial_5 C_0$, assuming this derivative acquires a vev.
The full set of supersymmetry-breaking terms,
which are all soft, can be summarized as
\begin{eqnarray} 
\label{soft}
&{\mathcal{L}}_{soft} = - \frac{2 V_0}{3 \kappa^2} \sum_{m=1,2}
\int_{M^{(m)}_{4}} d^4 x \sqrt{g^{(4)}_{m} } e^{-3 \sigma_m}
(g^{55 \, m})^3  \langle \partial_5 C_0 |_{m} \rangle \times &
\nonumber \\ & \left [ \frac{\sqrt{2} \kappa^2}{\lambda^2} P^{(m)}
- \frac{ 1}{64 \pi } \left ( \frac{\kappa}{4 \pi} \right )^{2/3}
e^{\frac{9}{2} \sigma_m} (g_{55 \, m})^{3/4}
  \bar{\epsilon}^{(4)}_{m} \epsilon^{(4)}_{m} \right ]&
\end{eqnarray} 
where the subscripts $m$ denote the values of the 
corresponding five-dimensional
fields on the $m$'th wall. 
Modulo certain details to be
discussed below, we see in (\ref{soft}) soft terms that are 
proportional to $P^{(m)}$ and trilinear in matter scalars,
which were unnoticed in early
papers, and soft gaugino masses, in both the hidden and the visible
sectors. Terms which are absent above include nonchiral
soft scalar masses. This absence of soft scalar masses is directly
related to the form of the bulk-wall coupling, i.e., to the form of
the perfect-square coupling, which was a consequence of
eleven-dimensional supersymmetry, in the order to which the
Horava-Witten Lagrangian was constructed.
The deeper reason for the absence of explicit
scalar masses is that we are only compactifying the
eleven-dimensional field
theory and not the full strongly-coupled string theory, i.e., we 
do not perform 
the equivalent of the orbifold compactification in the weakly-coupled
string, since the full theory is currently
inaccessible only in the strongly-coupled limit.

In order to see how the soft terms
found above would look in the presence of the non-universal
hypermultiplets in the bulk, i.e., for Calabi-Yau spaces with Hodge
number $h_{(2,1)} \, > \, 0$, we recall~\cite{elp} that
one has to make the replacement 
\beq
\label{repl} 2 \partial_5 C_0 \, \rightarrow \, \frac{2}{1 - |Z^a
|^2 } \left (
\partial_5 C_0 + \partial_5 C_a Z^a \right )
\eeq 
in the expression (\ref{soft}),
where $a=1,...,h_{(2,1)}$ labels $Z_2$-odd ($C_a $) and
$Z_2$-even ($Z^a$) scalars from the non-universal hypermultiplets.
The presence of the derivatives $\partial_5 C_a$ signals that, in
addition to the $F$ terms of the universal moduli $\tilde{S},\, \tilde{T}$ 
there
will be $F$ terms of the non-universal $(2,1)$ $\tilde{U}$ 
moduli participating in the
breaking of supersymmetry at low energy.

We now discuss the interpretation of the
solution (\ref{p5c}) from the point of view of the low-energy
supergravity Lagrangian. First of all, we note once more that, if we
switch off the gaugino bilinears, then what is left in $\partial_5
C_0$ is the scalar component of the superpotential, which happens
to be {\em the sum} of the superpotentials from individual walls.
This is exactly what we expect from the point of view of the 
four-dimensional supergravity, but it could not be taken for granted
from the beginning. Our procedure is the only way to recover all the
supergravity
terms involving the scalar component of the superpotential,
which we take as a clear sign that integrating out the fifth dimension,
as done here, is the canonical way of recovering the correct
four-dimensional effective theory. We now consider the effect of
switching on the gaugino condensates. It is clear
that they can be regarded 
as effectively non-dynamical parts of the superpotential
in the infrared limit. The 
consistent way to summarize these observations is to write 
\beq
\partial_5 C_0 = - \frac{1}{256 \pi^2 \rho_0} \left (
\frac{\kappa}{4 \pi} \right )^{2/3} \sum_m W^{(m)} \eeq 
where (using $g_{55}= e^{2 \gamma} $):
\beq
\label{sup} W^{(m)} = 64 \sqrt{2} P^{(m)} - 2 e^{\frac{3}{2}(3
\sigma_m + \gamma_m )} \langle \bar{\epsilon}^{(4)}_m
\epsilon^{(4)}_m \rangle 
\eeq 
is the
effective low-energy superpotential generated on the $m$'th wall.
This identification clearly can be done at the level of the 
lowest-order solution for $\partial_5 C_0$. There, the condensates
enter with exactly the same coefficients, and one can conclude that
the four-dimensional scalar potential obtained from the solution
(\ref{p5c}) is that 
the one of the four-dimensional supergravity with $f_{1,2} =
\tilde{S}$: this leads to
\beq
F^{\tilde{S}} \propto W^1 + W^2 - \frac{1}{4} ( \tilde{S} +
\bar{\tilde{S}})^{3/2} (\tilde{T} + \bar{\tilde{T}} )^{3/2} 
( \Lambda^{3}_1 + \Lambda^{3}_2 ),
\eeq
which agrees with (\ref{p5c}) and (\ref{sup}) at
the lowest order, i.e., with $\sigma_m \rightarrow \tilde{\sigma},
\; \gamma_m \rightarrow \tilde{\gamma}$. In this case,  we
immediately recover the original conclusion drawn in
eleven dimensions~\cite{hg}, namely that two condensates on opposite
walls with the same magnitudes and opposite phases result in
unbroken supersymmetry. This is in contrast to the result obtained
previously in four-dimensional supergravity with gauge kinetic functions
linear in $\xi_0$.
However, the rederivation of the result of~\cite{hg} in 
five dimensions on the
basis of (\ref{p5c}) in the latter case, with higher order corrections 
switched on, is misleading, since it is the dynamics which
must choose the physical configuration of fields from the family of all
those possible, which we have not yet fully taken into account.
We come back to an explanation of this point at the end of this
Section.

Actually, the above is not the only way 
in which supersymmetry breaking due to
condensates can be modified. One should note that the reduction
of the ten-dimensional Chern-Simons
forms (including the gravitational ones) can result not only in
trilinear terms $P^{(m)}$, but also in constant terms,
corresponding to non-zero fluxes of the gauge and tangent
vacuum bundles. These  terms, which are
constant from the four-dimensional point of view,
might also cancel condensates, causing the vev of
$\partial_5 C_0$ to vanish. To decide what is the actual vev
controlling supersymmetry breaking is a
dynamical problem, which can be addressed in the context of
the results of the present paper. Unfortunately, as predicted
already in~\cite{elp}, the effective potential which one obtains has a
runaway direction.

Lacking clearer options, we assume whenever
necessary that, if $\partial_5 C_0$ is non-zero, then it is of
the order of the effective condensation scale $\Lambda^3 / 2 \pi
\rho_0$. In this connection, we observe
on the basis of the exact
expression (\ref{soft}) that soft terms born on a
given wall are proportional to $\partial_5 C_0 |_m$. i.e., the
derivative of $C_0$ computed locally on that wall. There is no
five-dimensional averaging involved: soft terms are generated by
{\em local} interactions on the walls.

This observation is important, as it leads to the next consistency
check of the implicit assumption of the existence of the
effective four-dimensional supergravity. The obvious
place to read off the effective superpotential in
a given four-dimensional supergravity model is the gravitino mass term,
which is $e^{K/2} W$ in the canonical normalization for the
gravitino in four dimensions. When
one reduces the only eleven-dimensional term which can give rise to the
gravitino
mass term~\footnote{We focus on the highest-helicity components, as
the lower-helicity massive
components must come from the super-Higgs effect.} to lower
dimensions, one finds 
\beq 
-\frac{\sqrt{2} }{48 \kappa^2} \int d^{11} x
\sqrt{g} \bar{\psi}_{\mu} \Gamma^{\mu \nu A B C 11} \psi_{\nu}
G_{11 A B C} 
\eeq 
and so obtains as the mass-like coefficient in
four dimensions the integral 
\beq 
\langle e^{-\frac{3}{2} ( \gamma +
\sigma )}
\partial_5 C_0 \rangle = \frac{1}{\pi \rho_0} \int_{0}^{\pi \rho}
d x^5 e^{-\frac{3}{2} ( \gamma + \sigma )}
\partial_5 C_0. 
\eeq 
We recall first that $\frac{1}{4}
e^{-\frac{3}{2} ( \gamma + \sigma )}$ is indeed $e^{K/2}$. In order
to be consistent with the 
soft terms found previously, which are
proportional to the local values of the derivative $\partial_5
C_0$ on the respective walls, one must
demand that 
\beq 
\langle
\partial_5 C_0 \rangle = \partial_5 C_0 |_{(m=1)} =
\partial_5 C_0 |_{(m=2)} 
\eeq
One natural and  acceptable, but not of course unique,
solution to this condition is $\partial_5 C_0$ constant, which is
exactly the case for the solution of lowest order in $\kappa^{2/3}$
which we consider here. The obvious question is whether there is
any need to consider situations when the vacuum solution for $C_0$
is not exactly linear in $x^5$. The answer is that there exists a
natural source of such nonlinearities - the nonlinear couplings
from the non-trivial K\"ahler metric for the bulk moduli, i.e., the
non-trivial $\sigma$ model discussed in~\cite{elpp,elp}.
The incorporation of such nonlinear vacuum solutions into the
standard four-dimensional supergravity form lies beyond the scope of this 
paper: one may speculate that it could involve
field-theoretical loop corrections.

Finally, we would like to discuss in some detail the relation between the
five- and four-dimensional pictures of supersymmetry
breaking. To this end,
we first illustrate the way the corrections of higher order in
$\kappa^{2/3}$ differentiate between the terms coming from different walls.
We study the simplest
higher-order couplings related to Witten's deformation, in the operators
bilinear in gaugino fields.
The supergravity
Lagrangian with $f_{1,2}= \tilde{S} \pm \xi_0 \tilde{T}$ gives
\bea 
& \frac{e^{K/2}}{4} W g^{i \bar{j}} K_i
\partial_{\bar{j}} \bar{f}_{ab} \langle \bar{\lambda}^a
\bar{\lambda}^b \rangle \,+ \, h.c. \,=-\frac{e^{K/2}}{2} W \left
( \Lambda^{3}_1 (e^{3 \sigma } + \xi_0 e^{\gamma}) \right .&
\nonumber
\\ & \left . + \Lambda^{3}_2 (e^{3 \sigma } - \xi_0 e^{\gamma}) \right )
+ h.c.& 
\label{wasum} 
\eea 
where $\Lambda^{3}_i$ is the four-dimensional condensate on the
$i$'th wall. The first term on the right-hand side of (\ref{wasum}) looks
like the
contribution from the first wall, and the second 
like the one from the other wall.
Indeed, these terms are very similar to the two
expressions for the gaugino bilinears coming from different walls
which we obtained in (\ref{soft}), once we take into
account the linear correction to the volume: 
\beq
\sqrt{g_{6}(x^5)}
\rightarrow
\sqrt{g_{6}^{(0)} (x^5)} ( 1 + \xi_0 e^{\gamma} 
(1 - \frac{2 x^5 }{\pi \rho_0})),
\eeq
which expresses the difference between the volumes of the Calabi-Yau
spaces on
the two walls. Moreover, we note that $\partial_5 C_0 \propto W^{(1+2)}$ 
at the zeroth order in $\kappa^{2/3}$.

Pursuing explicitly this quest for higher-order operators
cannot give reliable results because of the
problems discussed above the equation (\ref{sigm5}).
However, we stress that the procedure of
reducing from five to four dimensions which we have applied
here already has many general
features of the exact reduction, and illustrates the
way the two spatially-separated gauge sectors are glued together.
This is the case not only in the Horava-Witten models
with spatially separated $9$-branes, but also, we believe, in more
general situations like those in Type-I/Type-IIB orientifold
models.

As the next step towards a better understanding of the relation
between the five- and four-dimensional  pictures,
we return to the four-dimensional expression for the auxiliary fields
controlling supersymmmetry breaking. We recall 
that the condition of unbroken 
supersymmetry $F^{\tilde{S}} = F^{\tilde{T}} = 0$ is
fullfilled if and only if both condensates vanish.
On the other hand, in general~\footnote{Namely, including
higher-order corrections in $\kappa^{2/3}$ (or $\xi_0$).} in
the five-dimensional picture: $\partial_5 C_0 = \alpha_1
\Lambda^{3}_1 + \alpha_2 \Lambda^{3}_2$, with $\alpha_1 = \alpha_2$
in the lowest-order solution. Hence, in five dimensions, unbroken
supersymmetry
does not imply $\Lambda^{3}_1=\Lambda^{3}_2=0$ as in four
dimensions. The resolution of the puzzle lies in the dynamics. One
has to remember that, among the possible values of the parameter that
controls supersymmetry breaking, the allowed ones are those which
extremize the energy functional. In the present case, the relevant
contributions to the energy are the gradient terms corresponding
to the derivatives with respect to $x^5$: 
\beq 
KE_5 = \int_{M^5}
\frac{2}{S + \bar{S}} ( 2 \partial_5 C_0 + ...)^2 + \int_{M^5}
\frac{1}{(S + \bar{S})^2} (\partial_5 S)^2 + ... \; .
\eeq 
After
integrating $C_0(x^5)$ and $S(x^5)$ out using their equations
of motion, these terms become potential-energy terms from the
point of view of four dimensions. Then,
setting $\partial_5 C_0 = 0$ does not extremize the full energy
functional. Since we use dynamical equations of motion
to integrate out non-zero mode parts of $C_0$ and
$S$, the information  that $\partial_5 C_0=0 $ is not their solution 
when one includes all the corrections, becomes encoded in the structure of
the
effective four-dimensional Lagrangian. In fact, the property that
not only the perfect square but also other derivatives with
respect to $x^{11}$ are important can be seen already in
eleven dimensions,
when one makes the Ansatz for the metric reflecting the canonical
choice of the four-dimensional moduli $\sigma$ and $\gamma$\footnote{In
the real
basis $x^i$, we now take for the metric on the internal six-dimensional 
space the metric $g_{ij} = \frac{1}{2} e^{\sigma} \delta_{ij}$.}. When
one expands the eleven-dimensional Einstein-Hilbert action in terms of
these variables, one obtains
\begin{eqnarray}
& -\frac{1}{2} \int_{M^{11}} d^{11} x \sqrt{g} R^{(11)} =
-\frac{1}{2} \int_{M^{11}} d^{11} x \sqrt{g_{(4)}} e^{-\gamma -2
\sigma} \left (
  - \frac{1}{2} e^{\gamma + 2 \sigma} (3 \gamma^\mu \gamma_\mu +
 9 \sigma^\mu \sigma_\mu + 2 \Box \gamma + 4 \Box \sigma)\right .&
  \nonumber \\  & \left . + e^{-\sigma} (6 \partial_{11} \gamma \partial_{11}
\gamma + 16
\partial_{11} \gamma \partial_{11} \sigma + 30 \partial_{11}
\sigma
 \partial_{11} \sigma -4 \partial_{11}^2 \gamma -10 \partial_{11}^2 \sigma )
 \right ) &
\end{eqnarray}
which contains not only four-dimensional kinetic terms but also terms
quadratic in derivatives with respect to $x^{11}$. 

It is the total sum of
both types of both these and the terms coming from the walls
should be extremized.
The fact that gradient energy is, in general, non-zero in 
the five-dimensional bulk is reflected in the non-vanishing values of the 
four dimensional $|F^S |^2$ and $|F^T |^2$ contributions 
to the energy density.

\section*{Conclusions}

We have obtained in this paper soft supersymmetry-breaking
operators in perturbative non-standard embeddings in the Horava-Witten
model of the effective low-energy field-theory limit of the
strongly-coupled heterotic string. The important feature that we have
studied in detail is that, in these models, charged  matter lives in
two sectors with different gauge kinetic functions on two
ten-dimensional walls which are spatially separated. We have also
considered the possibility that gauginos may
condense on both walls. We have also
studied how such two sectors combine with each other to form the
effective supergravity in four dimensions. We have shown how the
soft supersymmetry-breaking terms are born in such a 
five-dimensional picture, and shown that 
this picture is consistent, at the level of 
the lowest-order solution to
the equation of motion along the fifth dimension, with the
effective four-dimensional results. We have also shown that
integration over the fifth dimension gives
the structure of a
four-dimensional supergravity
with the gauge kinetic functions $f_{1,2} = S$
at the lowest order in $\kappa^{2/3}$. 
We have found non-universality of the
soft terms on different walls, which is due to sign
differences in the gauge kinetic functions and in the corrections
to the kinetic functions of charged scalars.

We have argued that mixed $F^S / F^T$, and perhaps even $F^S$-dominated,
supersymmetry breaking may occur
in the class of models discussed here, and, we believe, 
also in their counterparts in more general constructions with different
gauge sectors separated in higher dimensions.
We also believe that other considerations 
in this paper should be helpful
in more general situations, like type I/type IIB orientifold
models with gauge sectors located on different branes.\\

\vspace{0.5cm}

During the completion of this project, we learnt of the
forthcoming work of J.P. Derendinger and R. Sauser~\cite{talk},
which addresses the problem of constructing the effective 
low-energy effective supergravity in also in the presence of
five-branes in the eleven-dimensional bulk. We thank J.P. Derendinger for
very useful discussions explaining his results.\\

\vspace{0.5cm}

\noindent{\bf Acknowledgments}: The authors thank Witold Pokorski for
his collaboration in the early stages of this project. This work was
been supported
partially by the European Commission programs
ERBFMRX--CT96--0045 and CT96--0090. Z.L. and S.P. also acknowledge
support by the Polish Commitee for Scientific Research Grant 2
P03B 052 16 (99-2000) and by the M. Curie-Sklodowska Foundation.

\end{document}